\documentclass[aps,pra,reprint,groupedaddress]{revtex4-2}
\usepackage{amsmath}
\usepackage{graphicx}
\usepackage{bbm}
\usepackage[colorlinks,linkcolor=blue,anchorcolor=blue,urlcolor=blue,citecolor=blue]{hyperref}
\begin{document}
\title{Squeezed light generated with hyperradiance without nonlinearity}


\author{Jun Li}
\affiliation{MOE Key Laboratory of Advanced Micro-Structured Materials, School of Physics Science and Engineering, Tongji University, Shanghai, 200092, China}
\affiliation{Department of Physics and Astronomy, University of Manitoba, Winnipeg R3T 2N2, Canada}
\author{Chengjie Zhu}
\email[]{cjzhu@suda.edu.cn}
\affiliation{School of Physical Science and Technology, Soochow University, Suzhou 215006, China}
\affiliation{Collaborative Innovation Center of Light Manipulations and Applications, Shandong Normal University, Jinan 250358, China}
\author{Yaping Yang}
\email[]{yang\_yaping@tongji.edu.cn}
\affiliation{MOE Key Laboratory of Advanced Micro-Structured Materials, School of Physics Science and Engineering, Tongji University, Shanghai, 200092, China}


\date{\today}
\begin{abstract}
	We propose that the squeezed light accompanied by hyperradiance is induced by quantum interference in a linear system consisting of a high quality optical cavity and two coherently driven two-level qubits. When two qubits are placed at the crest and trough of the standing wave in the cavity respectively (i.e., they have the opposite coupling coefficient to the cavity), we show that squeezed light is generated in the hyperradiance regime under the conditions of strong coupling and weak driving. Simultaneously, the Klyshko's criterion alternates up and down at unity when the photon number is even or odd. Moreover, the orthogonal angles of the squeezed light can be controlled by adjusting the frequency detuning pressure between the driving field and the qubits. It can be implemented in a variety of quantum systems, including but not limited to two-level systems such as atoms, quantum dots in single-mode cavities.
\end{abstract}

	\maketitle
	
	Squeezed light is a prominent nonclassical source in quantum science (such as quantum nonlinear optics) and technology (such as quantum information processing and quantum metrology) because of its special photon number distribution and lower quantum noise in one orthogonal degree of freedom (in particular, phase and amplitude) \cite{Walls1983}. Typically, squeezed states of light expand the ultimate limit of quantum phase noise for interferometry beyond the fundamental boundary of coherent light so that it has been employed in the gravitational wave detectors LIGO \cite{Aasi2013} and GEO 600 \cite{PhysRevLett.110.181101} to enhance sensitivity. Recently, squeezed light has been proposed to enhance the interaction exponentially between light and matter \cite{Lemonde2016, PhysRevLett.120.093602, PhysRevLett.120.093601}, control quantum phase transition \cite{PhysRevLett.124.073602, PhysRevB.104.064423}, induce optical nonreciprocity \cite{PhysRevLett.128.083604} and achieve photon blockade \cite{PhysRevA.100.053857} and realize other novel quantum phenomena \cite{Prajapati:21, PhysRevResearch.4.013014}. 
	
	The generation of squeezed states usually requires a nonlinear optical process due to its nonlinear photon statistics. Squeezed light was first produced using atomic sodium as a nonlinear medium via four-wave mixing in 1985 \cite{PhysRevLett.55.2409} and was soon followed with experiments employing optical fibers \cite{PhysRevLett.57.691}, nonlinear crystals \cite{PhysRevLett.57.2520} and semiconductor laser \cite{PhysRevLett.58.1000}. After that, a variety of schemes and more substantial squeezing (up to 15 dB \cite{PhysRevLett.117.110801}) have been predicted theoretically and realized experimentally with the rapidly development of quantum technology. Among them, the optomechanical systems, which behaves as an effective nonlinear medium due to intensity-dependent cavity length, have been considered as better squeezing materials possessing some unique advantages such as enhancement of displacement sensitivity, independent of the optical frequency and miniaturization \cite{PhysRevA.49.1337, PhysRevA.49.4055, Brooks2012, Safavi-Naeini2013, PhysRevX.3.031012}. To sum up the above, high optical nonlinearity is the motive power of most squeezing states of light \cite{10.1002/9781119009719.ch5}.
	
	In this Letter, we propose a scheme to achieve squeezed states of optical field with hyperradiance in a linear coupled two-qubits-cavity system which has been considered to study collective radiation behavior\cite{Pleinert:17, PhysRevA.96.013839}, (unconventional) photon blockade\cite{PhysRevA.95.063842, PhysRevA.100.063817, Xia:22}, dipole blockade\cite{arxiv.2106.11268} and multiphoton nonclassical state\cite{PhysRevA.99.043814}. For the out-phase case (the opposite coupling coefficient to the cavity for two identical qubits), the levels with two-photon number difference are linked by external pumping and the coupling between two-level qubits and cavity due to quantum interference effects. Simultaneously, single-photon excitation with two-qubit ground state is forbidden by destructive interference. Thus, it is confirmed by the Wigner functions and Klyshko's criterion \cite{KLYSHKO19967, PhysRevLett.119.023602} distribution $K_n$ that the photon number distribution of the squeezed state can be formed under weak pumping without detuning. Further, the process is also accompanied by superradiation and hyperradiation behavior because of the presence of zero-eigenvalue states in the each photon number space of the dress struture.
	
\begin{figure}
		\includegraphics[width=\linewidth]{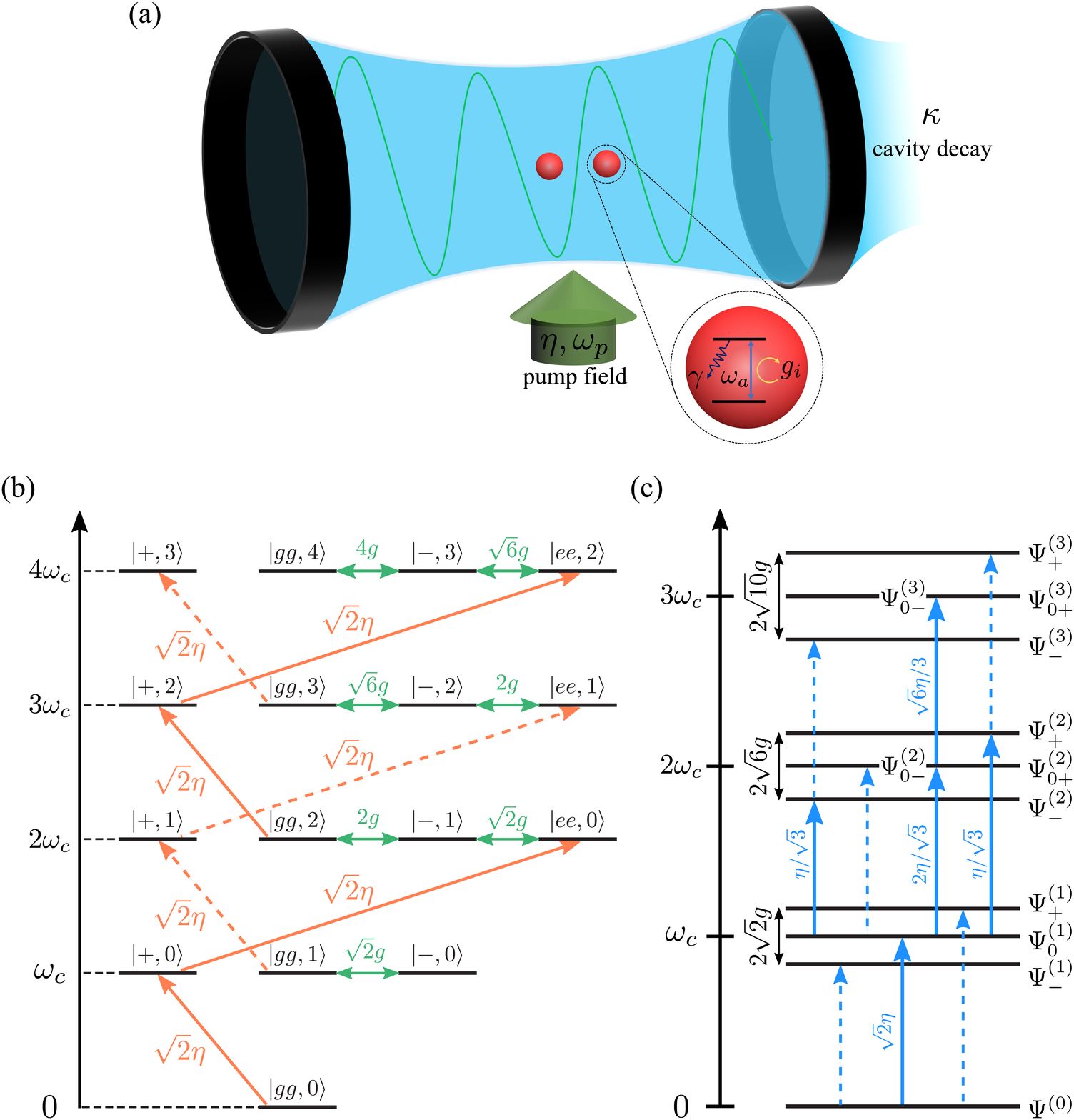}%
		\caption{\label{Fig.1}(a) The schematic illustration of two two-level systems (frequency $\omega_{a}$) driven by a coherent light of frequency $\omega_{p}$ coupled to a single-mode cavity (frequency $\omega_{c}$) with the coupling strength $g_i$. The cavity decay rate is $\kappa$ and the spontaneous emission rate of the excited state for both atoms is $\gamma$ respectively. (b) the energy level diagram and (c) the dressed-state structure of the out-phase coupling two-qubits-cavity system. Here, the collective states $|\pm\rangle= (|eg\rangle\pm |ge\rangle)/\sqrt{2}$ denote the symmetric and antisymmetric Dicke states. The single arrows in (b) and (c) denote the excitation pathways while the dashed arrows denote the forbidden transition pathways.}
	\end{figure}

	We consider a scheme consisting of two identical two-level qubits coupled to a high quality bosonic cavity as shown in Fig.\ref{Fig.1} (a). Analogous to the Jaynes–Cummings model, the dynamical behavior of the system can be described by the master equation
	
	\begin{equation}\label{eq:1}
		\dot{\rho}=-\frac{i}{\hbar}[H, \rho]+\kappa \mathcal{L} [a] \rho+\gamma \sum_{i=1,2} \mathcal{L} [\sigma^{i}]\rho,
	\end{equation}
	where $\rho$ is the density matrix operator of this two-qubits-cavity system. The Liouvillian superoperators $\mathcal{L}[a] \rho=2 a \rho a^{\dagger}-\rho a^{\dagger} a-a^{\dagger} a \rho$ and $\mathcal{L}[\sigma^{i}] \rho=2 \sigma_{-}^{i} \rho \sigma_{+}^{i} -\rho \sigma_{+}^{i} \sigma_{-}^{i}-\sigma_{-}^{i} \sigma_{+}^{i}\rho$ describe the losses of the system with the cavity decay at rate $\kappa$  and the spontaneous decay of the excited state of both qubits at rate $\gamma$, respectively. In a frame rotating with the frequency of the external field, the Hamiltonian of the two-qubits-cavity system can be expressed as
	
	\begin{equation}\label{eq:2}
		\begin{split}
			H/\hbar&=\Delta_{c} a^{\dagger} a+\Delta_{a}\sum_{i=1,2} \sigma_{+}^{i} \sigma_{-}^{i} \\
			&+\sum_{i=1,2}g_{i}\left(a^{\dagger}\sigma_{-}^{i}+a\sigma_{+}^{i}\right)+\eta\sum_{i=1,2}\left(\sigma_{-}^{i}+\sigma_{+}^{i}\right),
		\end{split}
	\end{equation}
	where $a^{\dagger}$ ($a$) is the bosonic creation (annihilation) operation of the cavity field. And $\Delta_{c}=\omega_{c}-\omega_{p}$ and $\Delta_{a}=\omega_{a}-\omega_{p}$ are the detunings with respect to the pump field frequency  $\omega_{p}$. Here, $\omega_{a}$ is the transition frequency of the qubits and $\omega_{c}$ is the resonance frequency of the cavity. $\sigma_{+}^{i}$ ($\sigma_{-}^{i}$) is the raising (lowering) operator of the $i$th ($i = 1,2$) two-level qubit. The third term represents the interaction between two atoms and the cavity, where $g_i$ is the coupling strength of the ith atom and the cavity. In the following, we consider the out-phase case (i.e., $g_1=-g_2=g$). The interaction between the two qubits and the monochromatic pumping field with Rabi frequency $\eta$ which is related to the input power. We just consider the simplest case where two qubits and the cavity have the same frequency, i.e., $\Delta_{a}=\Delta_{c}=\Delta$.
	
	Here, we introduce the collective states $|gg\rangle$, $|\pm\rangle= \left( |eg\rangle\pm |ge\rangle\right) /\sqrt{2}$ and $|ee\rangle$ of the two qubits as basis to express dressed state of the system clearly. The pumping and interaction Hamiltonian of the system can be rewrite as $\sqrt{2}\hbar \eta \left(D_{+}^{\dagger}+ D_{+}\right)$ and $\sqrt{2}\hbar g\left(a D_{-}^{\dagger}+a^{\dagger} D_{-}\right)$ in terms of the collective operators $D_{\pm}^{+}=\left(S_{1}^{+} \pm S_{2}^{+}\right) / \sqrt{2}$. In the absence of the pump field, we can obtain the eigenvalues $E_{\pm}^{n}=\pm\sqrt{4n-2}$ and $E_{0\pm}^{n}=0$ and the corresponding eigenstates $\Psi_{\pm}^{n}= \left(\sqrt{n} |g g, n\rangle \pm \sqrt{2n-1}|-, n-1\rangle+ |e e, n-2\rangle\right)/\sqrt{4n-2}$, $\Phi_{0}^{n}=\left( -\sqrt{n-1} |g g, n\rangle + |e e, n-2\rangle\right) / \sqrt{2n-1}$ and $\Psi_{0}^{n}=|+, n-1\rangle$ for multiphoton space ($n\geq2$), respectively. In one-photon space, the state $|ee\rangle$ does not exist, the eigenvalues are $E_{\pm}^{1}=\pm\sqrt{2}$ with the eigenstates $\Psi_{\pm}^{(1)}=(-|g g, 1\rangle \mp|-, 0\rangle) / \sqrt{2}$ and $E_{0}^{1}=0$ with the eigenstate $\Psi_{0}^{1}=|+, 1\rangle$. Notably, in every photon number space, the system always has a zero-eigenvalue state $\Psi_{0}^{n}=|+, n-1\rangle$. We plot the dressed-state structure and the corresponding excitation pathways while considering the coherent pump field in Fig.\ref{Fig.1} (c).
	
\begin{figure*}
		\includegraphics[width=\linewidth]{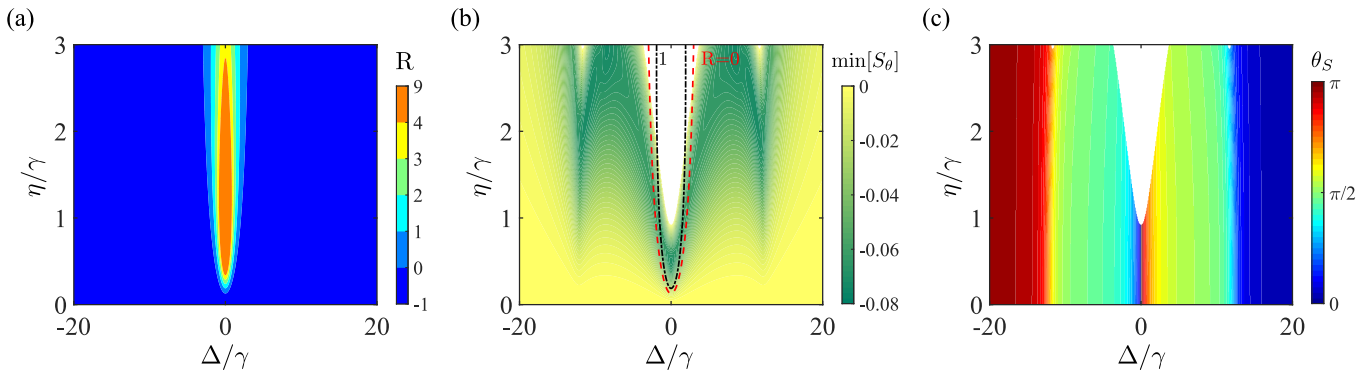}%
		\caption{\label{Fig.2}(a) The radiance witness $R$, (b) the minimum squeezing parameter min[$S_{\theta}$] and (c) squeezing angle $\theta_{S}$ mappings on the plane of the normalized detuning $\Delta/\gamma$ and driving strength $\eta/\gamma$ for the out-phase coupling two-qubits-cavity system. The blank areas in (b) and (c) indicate that the minimum squeezing parameter min[$S_{\theta}>0$]. The red dashed and black dot-dashed curves in (b) denote the contour of $R=0$ and $R=1$, respectively. Here, We set the coupling constant $g/\gamma = 10$ and the decay rate of cavity $\kappa/\gamma=0.5$.}
	\end{figure*}

	The collective radiance behaviors of the two-qubits-cavity system are quantified by the so called radiance witness $R$, which is defined as \cite{PhysRevA.88.063825, Pleinert:17}
	\begin{equation}\label{eq:3}
		R=\frac{\left\langle a^{\dagger} a\right\rangle_{2}-\sum_{i=1}^{2}\left\langle a^{\dagger} a\right\rangle_{1, i}}{\sum_{i=1}^{2}\left\langle a^{\dagger} a\right\rangle_{1, i}},
	\end{equation}
	where $\left\langle a^{\dagger} a\right\rangle_{2}$ and $\left\langle a^{\dagger} a\right\rangle_{1, i}$ are the average photon number for two qubits and the $i$th single qubit in the cavity, respectively. The radiance witness $R < 0$ indicates that the radiation is suppressed from the two qubits, corresponding to the so-called subradiance regime, while $R > 0$ reveals the radiation is enhanced. Particularly, the system is in the hyperradiance regime for $R > 1$. In Fig. \ref{Fig.2} (a), we plot the radiance witness $R$ of the out-phase coupling system as a function of the normalized detuning and driving strength by numerically solving Eq. (\ref{eq:1}) based on the Quantum Toolbox in MATLAB or QuTiP in PYTHON \cite{JOHANSSON20121760}. Near the resonant pumping ($\Delta=0$) area, the system exhibits a superradiant or hyperradiant behavior due to the multiphoton resonance excitation between the zero-eigenvalue states in the dress struture.
	
	Here, we consider the squeezing parameter $S_{\theta}$ to describe squeezing character of the cavity field by the width of the quadrature distribution as \cite{agarwal2012}
	\begin{equation}\label{eq:4}
		S_{\theta}=\left(\left\langle X_{\theta}^{2}\right\rangle-\left\langle X_{\theta}\right\rangle^{2}\right)-\frac{1}{2},
	\end{equation}
	where the general quadrature $X_{\theta}=(a e^{-\mathrm{i} \theta}+a^{\dagger} e^{\mathrm{i} \theta})/\sqrt{2}$ is a linear combination of the Hermitian quadrature operators $X=(a+a^{\dagger} )/\sqrt{2}$ and $Y=i(a^{\dagger} - a)/\sqrt{2}$. The $\theta\in [0,\pi)$ is the angle of the general quadrature relative to the quadrature operators $X$. It is obviously that the parameter has a lower bound $S_{\theta}\geq-1/2$. For a coherent state, the $S_{\theta}$ is always equal to 0 as the value of x changes. Any negative values of $S_{\theta}$ indicates the squeezing character of the field due to the narrower width of the quadrature distribution compared to coherent state. Moreover, the squeezing amplitude of the optical field is characterized by the minimum squeezing parameter min[$S_{\theta}$] among all quadrature directions and the corresponding angle $\theta_{S}$ is difined as squeezing angle.
	
	Additionally, to verify the nonclassicality of the photon number distribution, we introduce the Klyshko's criterion \cite{KLYSHKO19967} defined as
	\begin{equation}\label{eq:5}
		K_{n}=\frac{(n+1) P_{n-1} P_{n+1}}{n P_{n}^{2}} (n=1,2, \ldots),
	\end{equation}
	which can be obtained from the photon number distribution $P_{n}$ alone. The state is determined as nonclassical distribution if any value of $K_{n}$ is less than unity. Thus we have another sufficient condition for the nonclassicality of the field. For squeezed states, the photons are mainly distributed in the even number states of the Fock space while the odd number of photons are missing from. The photon number distribution of squeezed light are even-odd oscillation. Therefore, to some extent, the Klyshko's criterion with odd-even oscillation can reflect the squeezing property of the field \cite{PhysRevLett.119.023602}.
	
	In Fig. \ref{Fig.2} (b), we plot the the minimum squeezing parameter min[$S_{\theta}$] versus the normalized detuning $\Delta/\gamma$ and driving strength $\eta/\gamma$. Apparently, there are mainly three regions of squeezing parameter minima on the plane with the detunings $\Delta=0,\pm\sqrt{6} g/2$ for weak pumping ($\eta/\gamma<2$). For the case of off resonance at $\Delta=\pm\sqrt{6} g/2$, we have confirmed that two-photon excitations ($\Psi^{(0)} \rightarrow \Psi_{\pm}^{(2)}$) occur with three-photon blockade due to the nonlinearly dressed-state levels\cite{PhysRevA.95.063842}. The two-photon excitation process induces the generation of squeezed light. Interestingly, for the the case of resonant pumping $\Delta=0$, squeezed state of the cavity field is generated with superradiance or hyperradiance. In principle, as shown in Fig. \ref{Fig.1} (b), the transition path $|gg,n\rangle \stackrel {\sqrt{2}\eta} {\rightarrow} |+,n\rangle \stackrel{\sqrt{2}\eta}{\rightarrow} |ee,n\rangle \stackrel{g_{n1}}{\leftrightarrow} |-,n+1\rangle \stackrel{g_{n2}}{\leftrightarrow} |gg,n+2\rangle$ alway exists in each photon number space ($n\in N$) where $g_{n1}=\sqrt{2(n+1)}g$ and $g_{n2} = \sqrt{2(n+2)}g$. The states $|gg,n\rangle$ and $|gg,n+2 \rangle$ are connected through the above transition path while there is no connection between the adjacent photonic states $|gg,n\rangle$ and $|gg,n+1 \rangle$ for the system of ignored cavity decay. Furthermore, the state $|gg,n+1 \rangle$ cannot be excited because the two excitation paths $|gg,0\rangle \stackrel {\eta} {\rightarrow} |eg,0\rangle \stackrel{g}{\leftrightarrow} |gg,1\rangle$ and $|gg,0\rangle \stackrel {\eta} {\rightarrow} |ge,0\rangle \stackrel{-g}{\leftrightarrow} |gg,1\rangle$ interfere destructively. For the reason that the odd number states of photons for the two-qubit ground state $|gg\rangle$ and excited state $|ee\rangle$ have no photon population which is exactly the photon number distribution of squeezed state. Conversely, for the single-qubit excited states $|+\rangle$ and  $|-\rangle$, the excitation of even number states of photons is forbidden at the same time. As shown in Fig. \ref{Fig.2} (b), the squeezing amplitude of the optical field first increases to a maximum at $\eta=0.5\gamma$ and then decreases to vanish with the increase of strength of pump field while $\Delta=0$. It is because that the two-qubit state $|gg\rangle$ dominates for weak pumping and the single-qubit excited states $|+\rangle$ and  $|-\rangle$ dominate for strong pumping. The two mechanisms for producing squeezed light gradually merge with further enhancement of the driven field. The squeezed state of the cavity field exists in a wide range of detuning by controlling the strength of the driving field accordingly. In addition, the squeezing angle $\theta_{S}$ is almost entirely dependent on the detuning rather than the strength of the pump field as shown in Fig \ref{Fig.2} (c). So, the sqeezed light with any squeezing angle can also be achieved by adjusting the detuning as required such as amplitude squeezing and phase squeezing. Furthermore, the squeezing amplitude of the squeezed light can be further enhanced by providing more channels with more atoms.
	
	\begin{figure}
		\includegraphics[width=\linewidth]{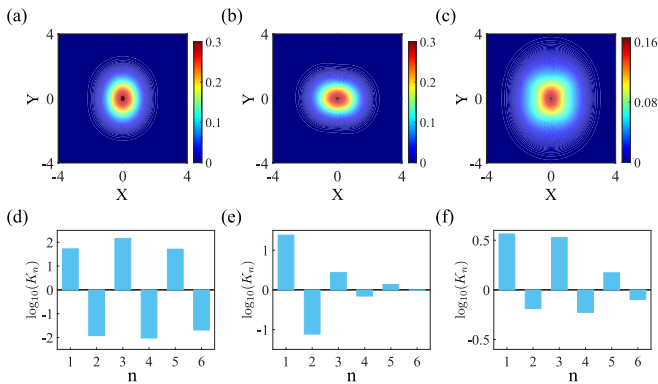}%
		\caption{\label{Fig.3} (a-c) The density plot of the Wigner functions and (d-f) corresponding Klyshko's criterion distributions $K_{n}$ for different points on the plane of Fig. \ref{Fig.2}. The strength and detuning of pump field are set as (a, d) $\eta/\gamma=0.5$ and $\Delta=0$, (b, e) $\eta/\gamma=2.65$ and $\Delta/\gamma=10$ and (c, f) $\eta/\gamma=2.5$ and $\Delta=0$, respectively. The other parameters of the system are the same as those used in Fig. \ref{Fig.2}.}
	\end{figure}
	
	On the one hand, we plot the Wigner functions of the cavity photon for three typecal points in Figs. \ref{Fig.3} (a-c) to show the probability distribution in phase space. In the squeezing regime, for Figs \ref{Fig.3} (a) and (b), the Wigner functions exhibit elliptical profile distribution indicating that fluctuations in a certain direction are suppressed compared to the larger range of fluctuation shown in Fig \ref{Fig.3} (c). It is obvious that they have almost vertical squeezing directions due to the different detuning of the pump field. On the other hand, we plot the corresponding Klyshko's criterion distribution $K_{n}$ of the cavity photon in Figs. \ref{Fig.3} (d-f) to show the nonclassical photon number distribution in Fock space. The Klyshko's criterion $K_n$ alternates up and down at unity denotes the even-odd photon number oscillation. For the resonant pumping $\Delta=0$, as shown in Fig. \ref{Fig.3} (d), The Klyshko's criterion $K_n$ maintains a large oscillation amplitude as $n$ increases. But for the case of off resonant pumping, $K_n$ oscillates greatly while $n=1, 2$ and decreases drastically with the increase of $n$. It reflects that the squeezed state of the cavity field in the off resonant regime is caused by the two-photon excitation process, while the squeezed light in the resonance regime can be induced by a higher-order even-numbered photon transition process. 
	
	In conclusion, unexpectedly from the previous work that demonstrated the existence of strong quantum fluctuations \cite{Pleinert:17}, we have shown the squeezed states of optical field generated with hyperradiance in an out-phase coupling two-qubits-cavity system without any nonlinearity. This arises from the quantum interference between the radiation fields of the two qubits. Moreover, the squeezed states also exist in case of off resonance pumping by the two-photon excitation process in the nonlinearly dressed-state levels. As the driving field increases, the squeezed states induced by the above two ways merge at the intermediate detuning. The squeezing angles of the squeezed light can be controlled by adjusting the detuning of driving field flexibly. Additionally, we also get the Wigner functions and Klyshko's criterion distributions as a verification and supplement to the above results.
	
	National Key Research and Development Program of China (2021YFA1400600 and No. 2021YFA1400602); National Nature Science Foundation (61975154 and No. 11874287); the China Scholarship Council (202106260079).
	
	\bibliography{Squeezedlight}

\end{document}